# Non-adiabatic holonomies as photonic quantum gates

Vera Neef[1], Julien Pinske[1], Tom A.W. Wolterink[1], Karo Becker[1],
Matthias Heinrich[1], Stefan Scheel[1], and Alexander Szameit[1]

[1]Institut für Physik, Universität Rostock, Albert-Einstein-Str. 23, 18059 Rostock, Germany
Contact: stefan.scheel@uni-rostock.de, alexander.szameit@uni-rostock.de

**Abstract:** One of the most promising nascent technologies, quantum computation faces a major challenge: The need for stable computational building blocks. We present the quantum-optical realization of non-adiabatic holonomies that can be used as single-qubit quantum gates. The hallmark topological protection of non-Abelian geometric phases reduces the need for quantum error correction on a fundamental physical level, while the inherent non-adiabaticity of the structures paves the way for unprecedented miniaturization. To demonstrate their versatility, we realize the Hadamard and Pauli-X gates, experimentally show their non-Abelian nature, and combine them into a single-qubit quantum algorithm, the *PQ penny flipover*. The planar geometry of such designs enables them to be substituted for the conventional directional coupler meshes currently in wide-spread use in photonic quantum architectures across all platforms.

Quantum technologies are on the rise. Exploiting properties that are inherent to, and unique in, quantum mechanics can dramatically improve the performance of well-established technologies such as data processing and storage, communication, security and sensing, while on the other hand opening up completely new areas of research ranging from quantum chemistry and biology to complex states of matter[1,2]. One key technology that has the capacity to revolutionize all of the above at once is the quantum computer. However, realizing such devices in a scalable, reliable and stable fashion has been proven a formidable challenge. Currently, several different technologies compete to be the future platform for quantum computers[3], including, but not limited to, superconducting circuits[4], trapped ions[5], and spins in solids[6]. Compared to their fermionic counterparts, photonic contestants[7–9], such as fiber loops or integrated waveguide circuits, bring some key advantages: They can be operated at room temperature, exhibit superior coherence times, and the small footprint of integrated optics promises both scalability and miniaturization. Yet, irrespective of the platform, the delicate nature of quantum systems necessitates extensive efforts in the field of quantum error correction to ensure the stability and reliability of computations. Here, we present a novel approach for the photonic implementation of single-qubit quantum gates leveraging non-adiabatic and non-Abelian holonomies as topologically protected and, thus, inherently more stable computational building blocks for integrated quantum-optical circuitry.

Quantum mechanics distinguishes between two types of phases: dynamical and geometric ones. While the former can be understood to record the passage of time, the latter depend exclusively on the geometry of the underlying Hilbert space and the path through it. A famous example of such a quantity is the scalar-valued Berry-Pancharatnam-phase[10,11], which is Abelian. In general, however, geometric phases can be non-Abelian, or non-commuting, and thus matrix-valued. They give rise to multidimensional unitary operators, so called holonomies[12], that describe the entire, purely geometric, evolution. Harnessing such holonomies as computational building blocks leads to the notion of holonomic quantum computing[13]. Notably, their geometric nature grants them topological protection[14,15], rendering these holonomic quantum gates inherently more stable and resilient against errors than their dynamical counterparts. The original proposition of holonomies by Wilczek and Zee[12] demands adiabaticity: The entire evolution is restricted to a degenerate eigenspace, i.e. a subspace of the Hilbert space with eigenenergy zero. This demanding requirement is the main source of error when using adiabatic holonomies as quantum gates, and the straightforward method to keep such deviations in check is to have the system vary as slowly as possible[16,17], imposing severe limitations on the degree of miniaturization and freedom in design. Yet, a purely geometric evolution can also be achieved without adiabaticity. By allowing propagation across a subspace containing states with a *mean* energy of zero, the geometric subspace $\mathcal{H}_{\text{geo}}$, and requiring that all dynamical contributions cancel out, the evolution is entirely described by a non-adiabatic holonomy[18] (Fig. 1). This approach substantially broadens the freedom of design while simultaneously enabling a much smaller footprint compared to adiabatic approaches without compromising topological error resilience[19] (see further: SI). While adiabatic holonomies have recently been implemented in photonic systems[16,17,20], their non-adiabatic extensions, currently are limited to other platforms such as: superconductors[21,22], nuclear magnetic resonance systems[23], solid-state spin systems[24], and trapped ions[25].

To construct a non-adiabatic holonomy, the geometric subspace $\mathcal{H}_{\text{geo}}$ of the Hilbert-space is spanned by the states $|\phi_l(z)\rangle$, which fulfil the non-adiabatic holonomic condition[18] $\langle \phi_m(z)|\widehat{H}(z)|\phi_l(z)\rangle = 0$ (with the Hamiltonian $\widehat{H}(z)$ and the propagation direction $z$). In a second-quantization formulation that allows for a complete description of a bosonic system, the states $|\phi_l(z)\rangle$ can be expressed as the excitations (action onto the vacuum state $|\mathbf{0}\rangle$) of the orthonormal modes $\widehat{\boldsymbol{\phi}}_k^\dagger(z)$. These modes fulfil the generalization of the non-adiabatic holonomic condition[25] $\left[\widehat{\boldsymbol{\phi}}_j, [\widehat{H}, \widehat{\boldsymbol{\phi}}_k^\dagger]\right] = 0$, ensuring a purely geometric evolution. Consequently, the mean energy of the Hamiltonian vanishes[26,27] in the geometric subspace. Our photonic system consists of three waveguides (denoted as left L, center C and right R). The planar geometry of our arrangement ensures that any direct coupling between the left and right channel is readily suppressed, creating the coupled mode Hamiltonian $\widehat{H}(z) = \kappa_{\text{L}}(z)\hat{a}_{\text{L}}\hat{a}_{\text{C}}^\dagger + \kappa_{\text{R}}(z)\hat{a}_{\text{R}}\hat{a}_{\text{C}}^\dagger + \text{h.c.}$, where $\kappa_{\text{L/R}}$ is the coupling coefficient between the left/right and central channel, $\hat{a}_\nu^{(\dagger)}$ represents the bosonic annihilation (creation) operator of the $\nu$th mode, and $z$ is the propagation direction. In this three-waveguide system, two modes $\widehat{\boldsymbol{\phi}}_1(z)$ and $\widehat{\boldsymbol{\phi}}_2(z)$ can be found that fulfill the non-adiabatic condition (see Methods). Placing a single photon in this system yields a two-dimensional geometric subspace $\mathcal{H}_{\text{geo}}$ spanned by $|\phi_1(z)\rangle = \widehat{\boldsymbol{\phi}}_1^\dagger(z)|\mathbf{0}\rangle$ and $|\phi_2(z)\rangle = \widehat{\boldsymbol{\phi}}_2^\dagger(z)|\mathbf{0}\rangle$. Imposing a cyclic evolution $|\phi_l(z_i)\rangle = |\phi_l(z_f)\rangle$, a non-adiabatic holonomy $\widehat{U}$ that is part of the two-dimensional unitary group U(2) exists such that it transforms any input state that is prepared in the initial geometric subspace $|\psi_{\text{in}}\rangle = c_1|\phi_1(z_i)\rangle + c_2|\phi_2(z_i)\rangle$ ($c_1, c_2 \in \mathbb{C}$) according to $|\psi_{\text{out}}\rangle = \widehat{U}|\psi_{\text{in}}\rangle$.

Following the parametrization [21,27] $\kappa_{\text{L}}(z) = \Omega(z)\sin\frac{\theta}{2}e^{i\varphi}$, $\kappa_{\text{R}}(z) = \Omega(z)\cos\frac{\theta}{2}$, the couplings are factorized into a constant weight (represented by $\theta$ and $\varphi$) and a common envelope function $\Omega(z)$ that contains all time- or $z$-dependency. Consequently, the cyclicity condition can be expressed as $\int_{z_i}^{z_f} \Omega(z)\text{d}z = \pi$, and, in turn, the holonomy transforms a qubit according to

$$\widehat{U}(\theta, \varphi) = \begin{pmatrix} \cos\theta & -e^{-i\varphi}\sin\theta \\ -e^{i\varphi}\sin\theta & -\cos\theta \end{pmatrix}, \quad (1)$$

with the qubit definition $|0\rangle_{\text{logic}} = |1_{\text{L}}0_{\text{R}}0_{\text{C}}\rangle$ and $|1\rangle_{\text{logic}} = |0_{\text{L}}1_{\text{R}}0_{\text{C}}\rangle$. As such, these logical states span the geometric subspace at the beginning and end of the propagation. Crucially, any intermediate state with non-zero detection probability of the photon in the central waveguide is not part of this initial geometric subspace. Evidently, a finite product sequence of holonomies $\widehat{U}(\theta, \varphi)$ suffices to realize any single-qubit transformation. To reduce experimental complexity, the couplings were chosen to be real and positive ($\varphi = 0$), resulting in a real, yet still non-Abelian holonomy. Moreover, the envelope function $\Omega(z)$ of any gate realization can be chosen freely, as long as it fulfills the cyclicity condition. This freedom in design, as illustrated in Fig. 2, constitutes a key advantage of non-adiabatic holonomic quantum gates as the waveguide trajectories can be chosen to simultaneously optimize for secondary goals such as the avoidance losses due to curvature or rapid changes in waveguide direction. Note that *adiabatic* holonomies constitute an inherently different class of holonomies, that require at least four waveguides to implement the same two-mode transformation[16].

We experimentally implement non-adiabatic holonomies in photonic waveguides, that are inscribed by femtosecond laser pulses into a fused silica chip[28]. A coupling scan allows for the easy translation between couplings and real-space distance (see Methods). We first demonstrate a holonomy that instantiates the (negative) Hadamard gate,

$$\hat{U}_\mathrm{H} = -\frac{1}{\sqrt{2}}\begin{pmatrix} 1 & 1 \\ 1 & -1 \end{pmatrix},$$

by choosing $\theta = 3\pi/4$. Here, the envelope function $\Omega(z)$ was chosen such that both left and right waveguide exhibit a straight section sandwiched between two cosine sections serving to separate the channels (Fig. 3 a). To allow for a smooth transition between the Hadamard gate and a standard fiber array both at the front and end facets, a cosine-shaped fanning section with negligible overall interactions was added to the outer waveguides. For our single-photon measurement, two identical photons are created through type-I spontaneous parametric down conversion, one of which serves as signal, and the other as a herald. By launching the signal photon into the left or right waveguide, respectively, and recording the coincidences between signal and herald, the absolute square of each matrix element of the holonomy can be measured, yielding an average fidelity (see Methods) between experiment and theory of $\bar{F} = (99.2 \pm 0.2)\%$. Here, the probability of the photon occurring in the central waveguide can be interpreted as a measure of how well the cyclicity condition is fulfilled. Only about 2.9% of all photons were detected in the central waveguide, when the left waveguide was chosen as the input, and about 0.3% for the right waveguide. Furthermore, this structure was employed as the beam splitter in a Hong-Ou-Mandel-type measurement[29], reaching a visibility of $(95.0 \pm 4.6)\%$ (see SI). Thus, the realized holonomy faithfully displays all properties of a Hadamard gate.

As a second demonstration, we realize a (negative) Pauli-X gate, also known as Flip or NOT gate,

$$\hat{U}_\mathrm{X} = -\begin{pmatrix} 0 & 1 \\ 1 & 0 \end{pmatrix},$$

a value of $\theta = \pi/2$ was chosen. Since $\sin\pi/4 = \cos\pi/4$, the Pauli-X gate is inherently symmetric: In line with their identical couplings, the left and right waveguide follow mirrored trajectories. The envelope function was chosen such that they follow a single cosine shape, and, due to the symmetry, no fanning section is needed. In the experiments, an average fidelity of $\bar{F} = (99.1 \pm 1.0)\%$ was obtained (Fig. 3 b). Together, the Pauli-X and Hadamard gates span the entire group of $2 \times 2$ unitaries U(2), which is associated with single-qubit computations.

To show that the realized holonomies are indeed non-Abelian, the commutator $\hat{U}_\mathrm{H}\hat{U}_\mathrm{X}\hat{U}_\mathrm{H} - \hat{U}_\mathrm{H}\hat{U}_\mathrm{H}\hat{U}_\mathrm{X}$ was examined by comparing the performance of the corresponding gate sequences Hadamard – Pauli-X – Hadamard and Pauli-X – Hadamard – Hadamard. (For more details see SI.) In this specific implementation, all gates consist of straight sections sandwiched between cosines, with intermediate fannings ensuring smooth transitions between subsequent differing gates. This new choice of geometry illustrates the freedom in design, granted by non-adiabatic holonomies. The stark difference in single-photon detection probabilities in Fig. 4 a) and 4 b) show clearly $\hat{U}_\mathrm{H}\hat{U}_\mathrm{X}\hat{U}_\mathrm{H} \neq \hat{U}_\mathrm{H}\hat{U}_\mathrm{H}\hat{U}_\mathrm{X}$ and, thus, the non-Abelian nature of the holonomies.

Finally, the non-adiabatic holonomic quantum gates were used to realize a quantum game: the *PQ penny flipover*[30,31]. In this game, P and Q take turns in manipulating the state of a "penny" in the sequence QPQ. Initially, the penny is in the state *heads* or $|0\rangle_\mathrm{logic}$. If the state after final measurement is *tails* or $|1\rangle_\mathrm{logic}$, P wins the game. While P is limited to the classical operations *flip* (i.e. employing the Pauli-X gate) or *no flip*, Q applies the Hadamard gate in both of their moves. As such, Q creates an eigenstate of the Pauli-X operator in their first move, ensuring that P's move has no effect and Q wins the game. The case where P flips the coin was realized as the gate sequence Hadamard – Pauli-X – Hadamard (Fig. 4b) and yields a $(99.0 \pm 1.2)\%$ chance for Q to win the game. In the case where P does not flip the coin, the Pauli-X gate was replaced by an inert section of decoupled waveguides (Fig. 4c), yielding a $(99.0 \pm 1.2)\%$ winning chance for Q. Thus, already this small holonomic algorithm shows an advantage of using a quantum game strategy, since no classical strategy could exceed a 50% chance of winning[30].

In our work, we have experimentally realized both a Pauli-X and Hadamard gate, employing non-adiabatic holonomies in a quantum-optical system, as well as short gate sequences that implement a quantum game and show the non-Abelian nature of the holonomies. This first demonstration of a non-adiabatic holonomic quantum gate sequence in integrated optics is an important step towards the photonic holonomic quantum computer. Implementing quantum gates and gate sequences with holonomies increases their stability as computational building blocks and reduces the need for error correction. Non-adiabatic holonomies, in particular, broaden the freedom of design while simultaneously enabling a much smaller footprint then their adiabatic counterparts. At the same time, they insulate the desired functionalities from inevitable fabrication inaccuracies such as local deviations in the waveguide separation or spectral walk-off with respect to the design wavelength, by linking the gate action exclusively to the cumulative interaction across the entire component. Crucially, existing photonic quantum technologies can be easily improved upon by replacing the currently ubiquitous directional coupler[9] in multi-qubit gates, algorithms or even reconfigurable photonic processors with non-adiabatic holonomies, thus, improving their stability and reducing the need for quantum error correction. Moreover, non-adiabatic holonomies will also prove useful beyond quantum computing: Only recently, the realization of an adiabatic holonomy belonging to the U(3) group was achieved[17]. Due to their less restrictive non-adiabatic holonomic condition, non-adiabatic holonomies are naturally of greater dimension then their adiabatic counterpart[26], allowing for the realization of holonomies of even greater dimension. For instance, a non-adiabatic U(3) holonomy can be implemented in a planar three-waveguide system without the need for any optimizations to reduce diabatic errors[26]. Consequently, the experimental study of gauge symmetries, quantum knots[32] and quantum braiding[33,34] could benefit from this platform. Similarly, the realization of non-adiabatic holonomies paves the way towards the realization of other types of holonomies in quantum optical systems, such as non-cyclic[35,36] and even non-Hermitian holonomies [37].

# References

1. Quantum Flagship. *EU* qt.eu.

2. National Quantum Initiative. *USA* quantum.gov.

3. Ladd, T. D. *et al.* Quantum computers. *Nature* **464**, 45–53 (2010).

4. Clarke, J. & Wilhelm, F. K. Superconducting quantum bits. *Nature* **453**, 1031–1042 (2008).

5. Cirac, J. I. & Zoller, P. Quantum computation with trapped ions. *Phys. Rev. Lett.* **74**, 4091–4094 (1995).

6. Loss, D. & DiVincenzo, D. P. Quantum computation with quantum dots. *Phys. Rev. A* **57**, 120–126 (1998).

7. Knill, E., Laflamme, R. & Milburn, G. J. A scheme for efficient quantum computation with linear optics. *Nature* **409**, 46–52 (2001).

8. Pelucchi, E. *et al.* The potential and global outlook of integrated photonics for quantum technologies. *Nat. Rev. Phys.* **4**, 194–208 (2022).

9. Wang, J., Sciarrino, F., Laing, A. & Thompson, M. G. Integrated photonic quantum technologies. *Nat. Photonics* **14**, 273–284 (2020).

10. Pancharatnam, S. Generalized theory of interference, and its applications - Part I. Coherent pencils. *Proc. Indian Acad. Sci. - Sect. A* **44**, 247–262 (1956).

11. Berry, M. V. Quantal phase factors accompanying adiabatic changes. *Proc. R. Soc. London* **392**, 45–57 (1984).

12. Wilczek, F. & Zee, A. Appearance of gauge structure in simple dynamical systems. *Phys. Rev. Lett.* **52**, 2111–2114 (1984).

13. Zanardi, P. & Rasetti, M. Holonomic quantum computation. *Phys. Lett. A* **264**, 94–99 (1999).

14. Solinas, P., Zanardi, P. & Zanghì, N. Robustness of non-Abelian holonomic quantum gates against parametric noise. *Phys. Rev. A - At. Mol. Opt. Phys.* **70**, 042316 (2004).

15. Parodi, D., Sassetti, M., Solinas, P., Zanardi, P. & Zanghì, N. Fidelity optimization for holonomic quantum gates in dissipative environments. *Phys. Rev. A - At. Mol. Opt. Phys.* **73**, 052304 (2006).

16. Kremer, M., Teuber, L., Szameit, A. & Scheel, S. Optimal design strategy for non-Abelian geometric phases using Abelian gauge fields based on quantum metric. *Phys. Rev. Res.* **1**, 033117 (2019).

17. Neef, V. *et al.* Three-dimensional non-Abelian quantum holonomy. *Nat. Phys.* **19**, 30–34 (2023).

18. Anandan, J. Non-adiabatic non-abelian geometric phase. *Phys. Lett. A* **133**, 171–175 (1988).

19. Johansson, M. *et al.* Robustness of nonadiabatic holonomic gates. *Phys. Rev. A - At. Mol. Opt. Phys.* **86**, 062322 (2012).

20. Yang, Y. *et al.* Synthesis and observation of non-abelian gauge fields in real space. *Sci.* **365**, 1021–1025 (2019).

21. Abdumalikov, A. A. *et al.* Experimental realization of non-Abelian non-adiabatic geometric gates. *Nature* **496**, 482–485 (2013).

22. Danilin, S., Vepsäläinen, A. & Paraoanu, G. S. Experimental state control by fast non-Abelian holonomic gates with a superconducting qutrit. *Phys. Scr.* **93**, 055101 (2018).

23. Feng, G., Xu, G. & Long, G. Experimental realization of nonadiabatic holonomic quantum computation. *Phys. Rev. Lett.* **110**, 190501 (2013).

24. Arroyo-Camejo, S., Lazariev, A., Hell, S. W. & Balasubramanian, G. Room temperature high-fidelity holonomic single-qubit gate on a solid-state spin. *Nat. Commun.* **5**, 4870 (2014).

25. Ai, M. Z. *et al.* Experimental Realization of Nonadiabatic Holonomic Single-Qubit Quantum Gates with Optimal Control in a Trapped Ion. *Phys. Rev. Appl.* **14**, 054062 (2020).

26. Pinske, J. & Scheel, S. Geometrically robust linear optics from non-Abelian geometric phases. *Phys. Rev. Res.* **4**, 1–10 (2022).

27. Sjöqvist, E. *et al.* Non-adiabatic holonomic quantum computation. *New J. Phys.* **14**, 103035 (2012).

28. Szameit, A. & Nolte, S. Discrete optics in femtosecond-laserwritten photonic structures. *J. Phys. B At. Mol. Opt. Phys.* **43**, 163001 (2010).

29. Hong, C. K., Ou, Z. Y. & Mandel, L. Measurement of Subpicosecond Time Intervals between Two Photons by Interference. *Phys. Rev. Lett.* **59**, 2044–2046 (1987).

30. Meyer, D. A. Quantum strategies. *Phys. Rev. Lett.* **82**, 1052–1055 (1999).

31. Müller, R. & Greinert, F. Playing with a Quantum Computer. *arXiv* **2108.06271**, (2021).

32. Xu, J. S. *et al.* Photonic implementation of Majorana-based Berry phases. *Sci. Adv.* **4**, 1–7 (2018).

33. Zhang, X. L. *et al.* Non-Abelian braiding on photonic chips. *Nat. Photonics* **16**, 390–395 (2022).

34. Iadecola, T., Schuster, T. & Chamon, C. Non-Abelian Braiding of Light. *Phys. Rev. Lett.* **117**, 073901 (2016).

35. Samuel, J. & Bhandari, R. General Setting for Berry's Phase. **60**, 2339–2342 (1988).

36. Mostafazadeh, A. Noncyclic geometric phase and its non-Abelian generalization. *J. Phys. A. Math. Gen.* **32**, 8157 (1999).

37. Garrison, J. C. & Wright, E. M. Complex geometrical phases for dissipative systems. *Phys. Lett. A* **128**, 177–181 (1988).

38. Brihaye, Y. & Kosiński, P. Adiabatic approximation and Berry's phase in the Heisenberg picture. *Phys. Lett. A* **195**, 296–300 (1994).

39. Nielsen, M. & Chuang, I. *Quantum Computation and Quantum Information*. (Cambridge Univ. Press, 2011).

## Methods

**Non-adiabatic holonomies**

The emergence of geometric phases can be readily described in the Heisenberg picture[38]. This formalism can be advantageous in the characterization of photonic networks, where it allows for a photon-number independent description. These ideas were formalized in Ref. [26] and apply to any linear optical multiport system. In particular, the disappearance of a Hamiltonian's dynamics $\hat{H}|_{\mathcal{H}_{\text{geo}}} = 0$ on a subspace $\mathcal{H}_{\text{geo}}$ is equivalent to the commutation relation $\left[\hat{\boldsymbol{\phi}}_j, [\hat{H}, \hat{\boldsymbol{\phi}}_k^\dagger]\right] = 0$. Here, the bosonic modes $\{\hat{\boldsymbol{\phi}}_k^\dagger(z)\}_k$ create excitations in the subspace $\mathcal{H}_{\text{geo}}$ through their action onto the vacuum $|0\rangle$. Given this condition, solutions to the Heisenberg equations of motion can be given explicitly[26] in terms of a holonomic superoperator $\hat{\mathcal{U}} = \mathcal{T} \exp\left(\int_{z_i}^{z_f} \mathcal{A}_z\right)$, where $\mathcal{T}$ denotes time-ordering and $(\mathcal{A}_z)_{jk} = \left[\hat{\boldsymbol{\phi}}_j, \partial_z \hat{\boldsymbol{\phi}}_k^\dagger\right]$ is an operator-valued generalization of the Anandan connection[18].

Let us apply the above construction to the three-waveguide coupler. Its coupling configuration reads $\kappa_L(z) = \Omega(z) \sin\frac{\theta}{2} e^{i\varphi}$, $\kappa_R(z) = \Omega(z) \cos\frac{\theta}{2}$. The two modes

$$\hat{d}^\dagger = \sin\frac{\theta}{2} e^{i\varphi} \hat{a}_R^\dagger - \cos\frac{\theta}{2} \hat{a}_L^\dagger,$$

$$\hat{b}^\dagger = \sin\frac{\theta}{2} e^{-i\varphi} \hat{a}_L^\dagger + \cos\frac{\theta}{2} \hat{a}_R^\dagger,$$

then satisfy $[\hat{H}, \hat{d}^\dagger] = 0$ and $[\hat{H}, \hat{b}^\dagger] = \hat{a}_C^\dagger$. Their evolution along the propagation length reads as[26]

$$\hat{\boldsymbol{\phi}}_1^\dagger(z) = \hat{d}^\dagger,$$

$$\hat{\boldsymbol{\phi}}_2^\dagger(z) = e^{i\delta(z)}\left(\cos(\delta(z)) \hat{b}^\dagger - i \sin(\delta(z)) \hat{a}_C^\dagger\right).$$

In other words, these modes satisfy the condition for a purely geometric propagation. Moreover, they evolve cyclically with $\delta(z_f) = \int_{z_i}^{z_f} \Omega(z) dz = \pi$. As such, the functionality of the gate only depends on structural parameters such was waveguide separation and component length as well as deviations from the design wavelength in so far as the cyclicity condition has to be fulfilled (see Supplementary Fig. S1). The connection is readily calculated to be $(\mathcal{A}_z)_{22} = i\Omega(z)$. The induced holonomy has a matrix representation $\mathcal{U} = \text{diag}(1, -1)$ with respect to the modes $\hat{\boldsymbol{\phi}}_1^\dagger$ and $\hat{\boldsymbol{\phi}}_2^\dagger$. Its action onto the spatial modes $\hat{a}_L^\dagger$ and $\hat{a}_R^\dagger$ is

$$\mathcal{U}(\theta, \varphi) = \begin{pmatrix} \cos\theta & -e^{-i\varphi} \sin\theta \\ -e^{i\varphi} \sin\theta & -\cos\theta \end{pmatrix}.$$

The operator holonomy $\mathcal{U}(\theta, \varphi)$ can create arbitrary mixing between the spatial modes. Hence, it can perform any two-mode transformation between the single photon states $\hat{a}_L^\dagger|0\rangle$ and $\hat{a}_R^\dagger|0\rangle$. The calculation for higher photon numbers is equivalent.

**Experimental configuration**

The waveguide structures used in our proof-of-principle experiments are inscribed into the bulk of fused silica chips (Corning 7980, dimensions 1 mm × 20 mm × 100 mm or 1 mm × 20 mm × 150 mm, bulk refractive index $n_0 \approx 1.453$ at 815 nm) by focusing ultrashort laser pulses from a frequency-doubled fiber amplifier system (Coherent Monaco, wavelength 517 nm, repetition rate 333 kHz, pulse duration 270 fs). In the focal volume, a localized and permanent refractive index change ($\Delta n_0 \approx 2 \times 10^{-3}$) is induced by nonlinear absorption and subsequent rapid quenching of the intermittent plasma back to room temperature. The desired waveguide trajectory is defined by the motion of a precision translation system (Aerotech ALS130, operated at: writing speed 100 mm min$^{-1}$), resulting in high-quality single-mode waveguides with propagation losses of approximately 0.3 dB/cm.

The coupling $\kappa$ between interacting waveguides is seamlessly controlled through the real-space distance $\Delta$ and follows an empirical exponential fit $\kappa = a \exp(-b\Delta)$. The experimental parameters exhibit typical values of $a \approx 20 \text{cm}^{-1}$ and $b \approx 0.2 \mu\text{m}^{-1}$, as characterized by a coupling scan, consisting of a set of directional couplers with increasing distance $\Delta$.

For our heralded single-photon measurements, pairs of indistinguishable photons are created via type-I spontaneous parametric down conversion in a bismuth borate ($BiB_3O_6$) crystal pumped with a 407 nm continuous-wave laser of 100 mW power. The signal photon is collected by a polarization-maintaining fiber, launched into the sample via a fiber array (pitch 82 μm), and subsequently collected by an array of multimode fibers that route it to avalanche photodetectors (Excelitas, detection efficiency of >50%, dark counts of <50 s$^{-1}$, dead time of 20 ns). Note that the intrinsic birefringence of laser-written waveguides ensures a preservation of the polarization degree of freedom throughout the entire chip. Meanwhile, the photon serving as herald is routed directly to a detector via a multimode delay fiber. Coincidences are retrieved with a correlation card (Becker & Hickl, time bin 164 ps). Here, a signal-herald coincidence is counted if both photons are detected within 5 ns of one another.

To compare the experimental results to the ideal gates or gate sequences, the average fidelity was calculated. The fidelity is defined as $F = |\langle \psi_{\text{out}}^{\text{theo}} | \psi_{\text{out}}^{\text{exp}} \rangle|^2$, where $|\psi_{\text{out}}^{\text{theo/exp}}\rangle$ denote the theoretical and experimental output states, respectively. Since the states $|0\rangle_{\text{logic}}$ and $|1\rangle_{\text{logic}}$ of the qubit are orthogonal, the output states can be expanded as $|\psi_{\text{out}}^{\text{theo/exp}}\rangle = \sqrt{p_{k0}^{\text{theo/exp}}}|0\rangle_{\text{logic}} + \sqrt{p_{k1}^{\text{theo/exp}}}|1\rangle_{\text{logic}}$, where $p_{kj}^{\text{theo/exp}}$ denotes the probability of the state $|j\rangle_{\text{logic}}$ being detected after $|k\rangle_{\text{logic}}$ was launched into the system ($k,j = 0,1$). Averaging over both input states yields the average fidelity $\bar{F} = \frac{1}{2}\sum_{j=0}^{1}\left(\sum_{k=0}^{1}\sqrt{p_{kj}^{\text{theo}} p_{kj}^{\text{exp}}}\right)^2$, also known as the Bhattacharyya distance[39]. The uncertainty of the mean fidelity $\bar{F}$ follows from the uncertainty of $p_{kj}^{\text{exp}}$, which was estimated as the sum of two major contributors: a single standard deviation of the counts for the Poisson noise and 1% of the total counts to account for variations in outcoupling efficiency.

**Acknowledgements**

We thank C. Otto for preparing the high-quality fused silica samples used for the inscription of all photonic structures employed in this work. A.S. acknowledges funding from the Deutsche Forschungsgemeinschaft (grants SZ 276/9-2, SZ 276/19-1, SZ 276/20-1, BL 574/13-1 and SZ 276/21-1), as well as the Krupp von Bohlen and Halbach Foundation. S.S. acknowledges funding from the Deutsche Forschungsgemeinschaft (grant SCHE 612/6-1). A.S. also acknowledges funding from the FET Open Grant EPIQUS (grant no. 899368) within the framework of the European H2020 programme for Excellent Science. T.A.W.W. is supported by a Marie Skłodowska-Curie fellowship from the European Commission (project no. 895254). A.S., S.S. and M.H. acknowledge funding from the Deutsche Forschungsgemeinschaft via SFB 1477 'Light–Matter Interactions at Interfaces' (project no. 441234705).

**Author Contributions**

V.N. designed and fabricated the waveguide structures and performed the measurements. J.P. provided the theoretical framework. V.N. and T.A.W.W. performed the stability analysis and investigated the commutating behavior. T.A.W.W. and K.B. helped in overcoming experimental challenges. M.H., S.S., and A.S. supervised the project. All authors co-wrote the manuscript.

**Competing Interests**

The authors declare no competing interests.

**Data availability**

Source data are provided with this paper in the Rostock University Publication Server repository (#####). All other data that support the plots within this paper and the other findings of this study are available from the corresponding authors upon reasonable request.

Correspondence and requests for materials should be addressed to Stefan Scheel or Alexander Szameit.

Figures

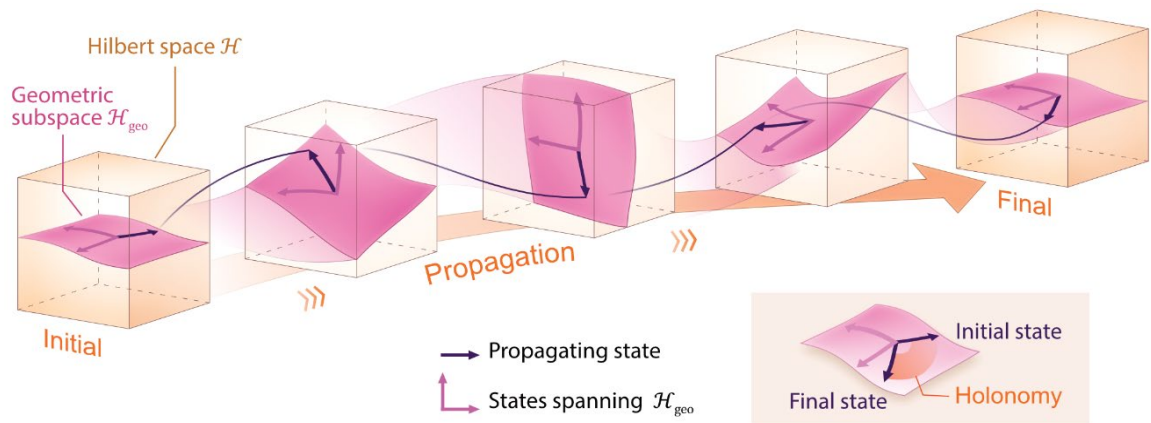

**Figure 1: Schematic depiction of a non-adiabatic holonomy:** The states spanning the geometric subspace $\mathcal{H}_{geo}$ change during the evolution of the quantum system. Consequently, $\mathcal{H}_{geo}$ itself changes during the propagation. Crucially, $\mathcal{H}_{geo}$ at the end of the propagation is identical to $\mathcal{H}_{geo}$ at the beginning. Any input state, prepared inside the initial $\mathcal{H}_{geo}$, will stay inside $\mathcal{H}_{geo}$ during its entire evolution. However, due to the non-trivial topology of $\mathcal{H}_{geo}$, the output state can differ from the input state. This transformation from input state to output state is called the holonomy (inset).

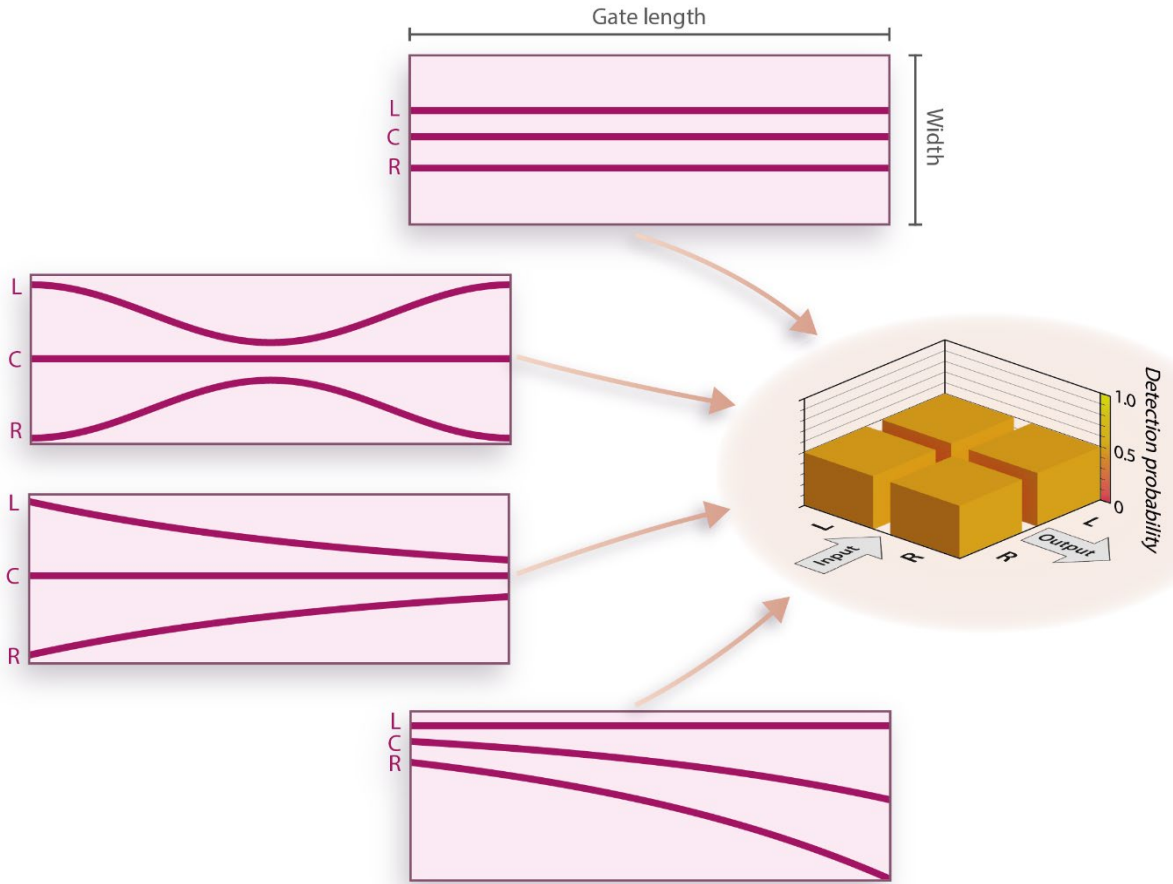

**Figure 2: Freedom of design.** The implementation of quantum gates based on non-adiabatic holonomies enables a wide variety of design choices, particularly in the choice of the envelope function $\Omega(z)$. As an example, four different yet functionally equivalent waveguide trajectories (left) that all instantiate a Hadamard gate are shown alongside their common distribution of detection probabilities (right).

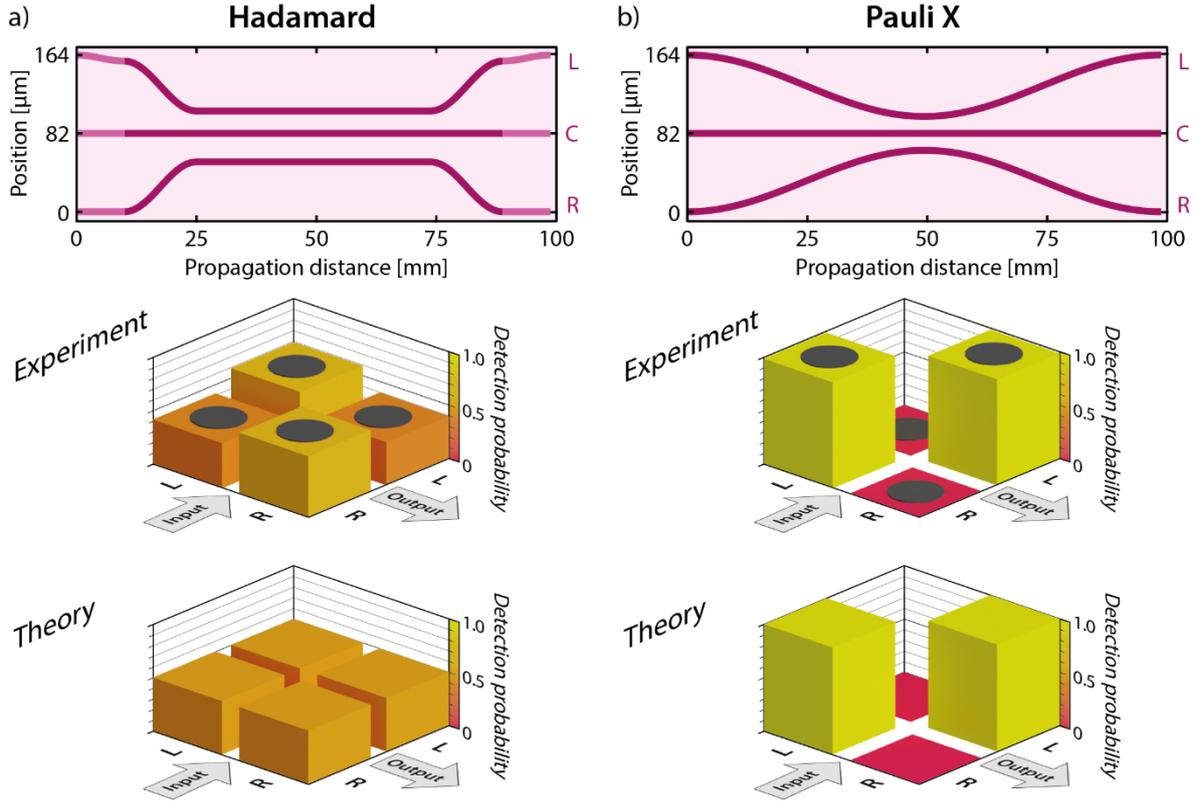

**Figure 3: Realization of two non-adiabatic holonomic single-qubit quantum gates. a)** Hadamard gate: The waveguide structure (top) consists of three waveguides in a planar arrangement. The trajectories of both the L and R waveguide consist of a straight section in between two cosine sections. At the front and end facets, fanning sections (indicated by lighter color) allow for a smooth transition towards standard fiber arrays. The experimental results (middle) show the probability of the photon being detected in L or R after being injected into either waveguide. The error bars (grey cylinders) are based on a single Poisson standard deviation as well as an estimation of the uncertainty of the outcoupling efficiency. The comparison with the theoretical expectations of the probability distribution (bottom) yields an average fidelity of $\bar{F} = (99.2 \pm 0.2)\%$. **b)** Pauli-X gate: Here, both L and R waveguide follow cosinusoidal trajectories (top). Due to the inherent symmetry of the gate, no fanning sections are needed. The experimental detection probability (middle) yields an average fidelity of $\bar{F} = (99.1 \pm 1.0)\%$.

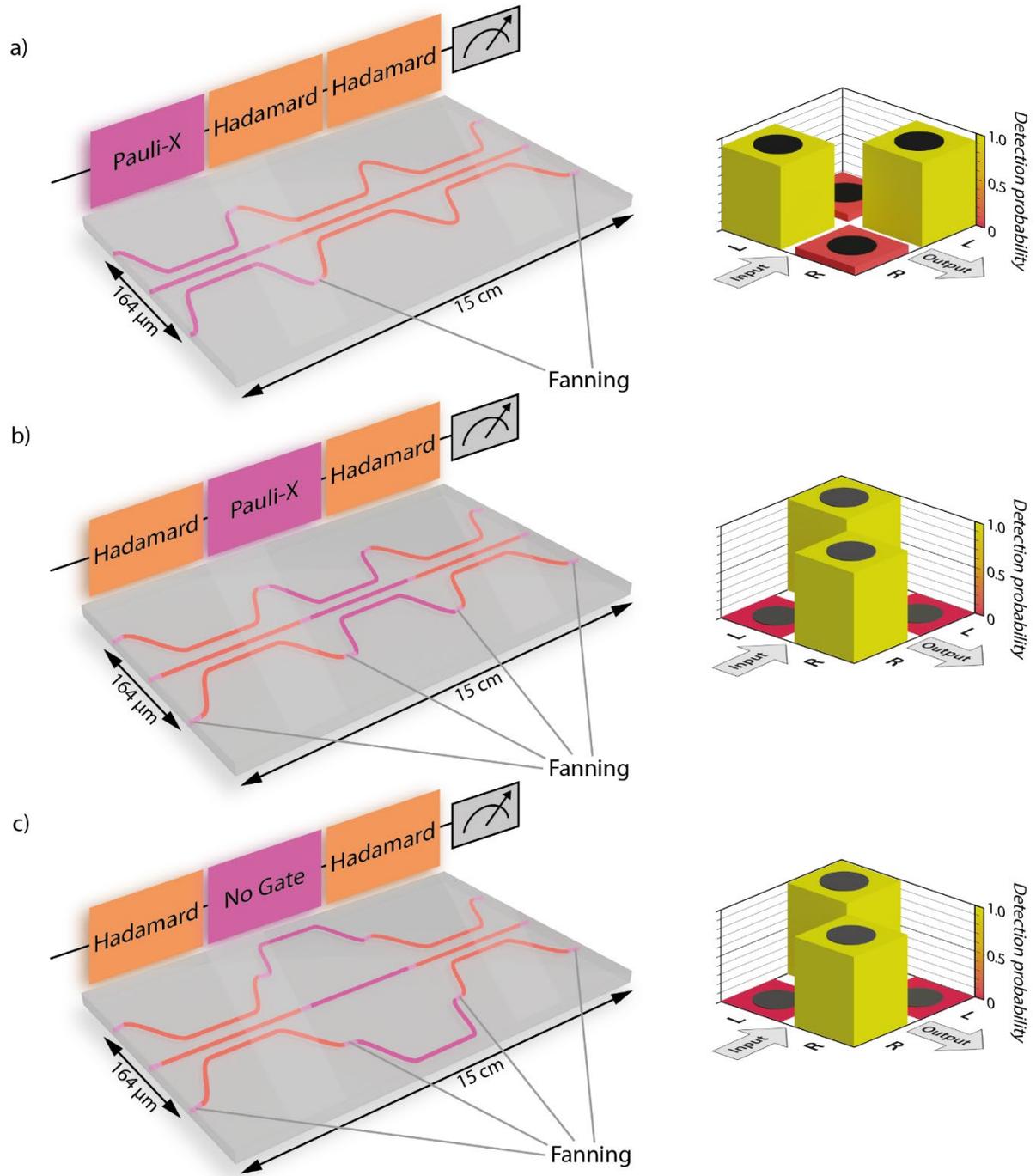

**Figure 4: Realization of gate sequences with non-adiabatic holonomic quantum gates.** Each individual gate consists of a section with straight waveguides in-between two sections of cosine-shaped outer waveguides. **a)** Pauli-X – Hadamard – Hadamard: Here fanning sections are added between the Pauli-X and Hadamard gate as well as at the end facet to allow for smooth transitions. **b)** Hadamard – Pauli-X – Hadamard: Here fanning sections are needed between any two gates as well as at the front and end facet. Comparing the experimental detection probabilities of **a)** and **b)** (right panel) shows clearly that the Hadamard gate does not commute with the sequence Pauli-X – Hadamard. **c)** Hadamard – no gate – Hadamard: In the *PQ penny flipover* [30], a quantum game, P can choose to either flip the penny or not flip the penny, while Q always applies a Hadamard gate. The case, where P does not flip the penny, is realized with a section, where the waveguides are completely decoupled. Here, both outer

waveguides mirror their respective trajectories in the Pauli-X gate in **b)** (where the penny is flipped over), furthermore both Hadamard gates and all fanning sections are identical in **b)** and **c)** to keep the two structures as similar as possible. The experimental detection probabilities (right panel) show clearly that whether the penny is flipped over or not does not change the outcome of the game.